\documentclass[aps,prb,twocolumn,showpacs,psfig,superscriptaddress,longbibliography]{revtex4-1}
\usepackage{amsfonts}
\usepackage{amssymb}
\usepackage{mathrsfs}
\usepackage{amsmath}
\usepackage{color}
\usepackage{natbib}
\usepackage{textcomp}
\usepackage{graphicx}
\usepackage{bm}
\usepackage{amssymb}
\usepackage{subfigure}
\usepackage{algorithm}

\usepackage{algorithmic}
\usepackage{xspace}
\usepackage{epstopdf}
\usepackage{dcolumn}
\usepackage{longtable}
\usepackage{makecell}
\usepackage{multirow}
\usepackage[colorlinks=true, letterpaper=true, pdfstartview=FitV, linkcolor=blue, citecolor=blue, urlcolor=blue]{hyperref}
\usepackage{float}
\usepackage{lipsum}

\makeatletter

\newcommand{\Rmnum}[1]{\expandafter\@slowromancap\romannumeral #1@}
\newcommand{\mycite}[1]{\scalebox{1.6}[1.6]{\raisebox{-0.9ex}{\cite{#1}}}}
\makeatother

\begin{document}

\title{Microscopic mechanism of high temperature ferromagnetism in Fe, Mn, Cr-doped InSb, InAs and GaSb magnetic semiconductors}

 \author{Jing-Yang You}
 \affiliation{Kavli Institute for Theoretical Sciences, and CAS Center for Excellence in Topological Quantum Computation, University of Chinese Academy of Sciences, Beijng 100190, China}

 \author{Bo Gu}
 \email{gubo@ucas.ac.cn}
 \affiliation{Kavli Institute for Theoretical Sciences, and CAS Center for Excellence in Topological Quantum Computation, University of Chinese Academy of Sciences, Beijng 100190, China}
\affiliation{Physical Science Laboratory, Huairou National Comprehensive Science Center, Beijing 101400, China}

 \author{Sadamichi Maekawa}
  \affiliation{Kavli Institute for Theoretical Sciences, and CAS Center for Excellence in Topological Quantum Computation, University of Chinese Academy of Sciences, Beijng 100190, China}
\affiliation{Center for Emergent Matter Science, RIKEN, Wako 351-0198, Japan}

 \author{Gang Su}
 \email{gsu@ucas.ac.cn}
 \affiliation{Kavli Institute for Theoretical Sciences, and CAS Center for Excellence in Topological Quantum Computation, University of Chinese Academy of Sciences, Beijng 100190, China}
 \affiliation{Physical Science Laboratory, Huairou National Comprehensive Science Center, Beijing 101400, China}
 \affiliation{School of Physical Sciences, University of Chinese Academy of Sciences, Beijing 100049, China}

\begin{abstract}
In recent experiments, high Curie temperatures Tc above room temperature were reported in ferromagnetic semiconductors Fe-doped GaSb and InSb, while low Tc between 20K to 90K were observed in some other semiconductors with the same crystal structure, including Fe-doped InAs, and Mn-doped GaSb, InSb, and InAs. Here we study systematically the origin of high temperature ferromagnetism in Fe, Mn, Cr-doped GaSb, InSb, and InAs magnetic semiconductors by combining the methods of density functional theory and quantum Monte Carlo. In the diluted impurity limit, the calculations show that the impurities Fe, Mn, and Cr have similar magnetic correlations in the same semiconductors. Our results suggest that high (low) Tc obtained in these experiments mainly comes from high (low) impurity concentrations. In addition, our calculations predict the ferromagnetic semiconductors of Cr-doped InSb, InAs, and GaSb that may have possibly high Tc. Our results show that the origin of high Tc in (Ga,Fe)Sb and (In,Fe)Sb is not due to the carrier induced mechanism because Fe$^{3+}$ does not introduce carriers. 
\end{abstract}
\pacs{}
\maketitle

\section{Introduction}
Magnetic semiconductors, which combine the dual characteristics of ferromagnets and semiconductors, are basic materials for next-generation semiconductor devices. Practical applications require that magnetic semiconductors can work at room temperature, which is a big challenge in science~\cite{Kennedy2005}. Over past three decades, magnetic semiconductor studies were focused on the diluted magnetic semiconductors. The representative material is Mn-doped p-type semiconductors (Ga,Mn)As~\cite{Ohno1998}, in which the highest Curie temperature Tc is currently about 200K~\cite{Chen2011}. How to increase Tc in (Ga,Mn)As has become a hot topic under intensive exploration~\cite{Dietl2010}. In order to avoid various difficulties caused by valence state mismatch in magnetic doping in (Ga,Mn)As, great progress has been made in the diluted magnetic semiconductors with  independently adjustable magnetic moments and carriers concentrations. In 2007, a theoretical study predicted the ferromagnetic semiconductor Li(Zn,Mn)As with decoupled magnetic moments and carriers doping~\cite{Masek2007}. In 2011, the experiment obtained the Mn-doped p-type ferromagnetic semiconductor Li(Zn, Mn)As with Tc of 50K~\cite{Deng2011}. In the same ferromagnetic semiconductor family, experiments realized Mn-doped p-type Li(Zn,Mn)P with Tc of 34K~\cite{Deng2011} and Li(Cd,Mn)P with Tc of 45K~\cite{Deng2013}.
In 2013, the experiment discovered Mn-doped p-type ferromagnetic semiconductor (Ba,K)(Zn,Mn)$_2$As$_2$ with a higher Tc of 230K~\cite{Zhao2013,Zhao2014}. Motivated by the experimental high Tc, the density functional theory calculation~\cite{Glasbrenner2014} and photoemission spectroscopy experiments~\cite{Suzuki2015,Suzuki2015a} were conducted to understand the microscopic mechanism of ferromagnetism in (Ba,K)(Zn,Mn)$_2$As$_2$. Since 2016, the Tc higher than room temperature has been reported in the experiments of Fe-doped p-type ferromagnetic semiconductor (Ga,Fe)Sb~\cite{Tu2016,Goel2019,Goel2019a,Takiguchi2019}. 

In addition to the p-type diluted ferromagnetic semiconductors, studies on the n-type diluted ferromagnetic semiconductors have also made great progress. In 2016, a theory proposed a physical picture of the diluted ferromagnetic semiconductors with narrow band gaps, which can form ferromagnetism controlled by n-type and p-type carriers, while diluted ferromagnetic semiconductors with wide band gaps can only form ferromagnetism controlled by p-type carriers~\cite{Gu2016,Gu2016a,Gu2019}. In 2017, the experiment reported the Tc higher than room temperature in Fe-doped n-type ferromagnetic semiconductor (In,Fe)Sb~\cite{Kudrin2017}, and later the experiment found that a gate voltage in (In,Fe)Sb can turn its Tc~\cite{Tu2018,Tu2019}. The origin of high Tc in (In,Fe)Sb and (Ga,Fe)Sb has been discussed based on the density functional theory calculations~\cite{Zhang2019}. In 2019, the experiment obtained a Tc of 45K in Co-doped n-type semiconductor Ba(Zn,Co)$_2$As$_2$~\cite{Guo2019}.

In contrast to the high Tc above room temperature obtained in Fe-doped p-type (Ga,Fe)Sb and n-type (In,Fe)Sb, low Curie temperatures of 25K and 10K were obtained in the experiments of Mn-doped p-type (Ga,Mn)Sb~\cite{Abe2000} and p-type (In,Mn)Sb~\cite{Ganesan2008}, respectively. In addition, Tc of 70K
was obtained in Fe-doped n-type (In,Fe)As~\cite{Hai2012,Hai2012a}, and 90K was obtained in the Mn-doped p-type (In,Mn)As~\cite{Schallenberg2006}.

In order to understand the microscopic mechanism of high Tc in Fe-doped GaSb and InSb, and low Tc in Fe-doped InAs and Mn-doped GaSb, InSb and InAs, in this paper we carry out a systematic theoretical study based on the combined method of density functional theory and quantum Monte Carlo. In the diluted impurity limit, our calculations show that the impurities Fe, Mn, and Cr have similar magnetic correlations in the same semiconductors. Thus, we argue that high (low) Tc in these experiments mainly comes from high (low) impurity concentrations. Our calculations predict ferromagnetic semiconductors Cr-doped InSb, InAs, and GaSb that may have high Tc. Our results show that the origin of high Tc in (Ga,Fe)Sb and (In,Fe)Sb is not due to the carrier induced mechanism because Fe$^{3+}$ does not introduce carriers. Thus, in order to increase Tc in diluted ferromagnetic semiconductors, a primary strategy is to increase the impurity concentrations while keeping the crystal structures unchanged, for example by choosing proper impurities and host semiconductors to avoid valence state mismatch during host doping.

\section{DFT+QMC METHOD}
We use a combination of the density functional theory (DFT)~\cite{Hohenberg1964,Kohn1965} and the Hirsch-Fye quantum Monte Carlo (QMC) simulation~\cite{Hirsch1986}. Our combined DFT+QMC method can be used for an in-depth treatment of band structures of host materials and strong electron correlations of magnetic impurities on equal footing. It can be applied for designing functional semiconductor-~\cite{Gu2008,Ohe2009,Gu2009} and metal-based~\cite{Gu2010,Gu2010a,Xu2015} materials. The method involves two calculations steps. First, with the Anderson impurity model~\cite{Haldane1976}, the host band structure and impurity-host mixing are calculated using the DFT method. Second, magnetic correlations based on the Anderson impurity model at finite temperatures are calculated using Hirsch-Fye QMC technique~\cite{Hirsch1986}. Considering the cost of calculation and the accuracy of the results, we set the temperature at 360K.

The Anderson impurity model is defined as follows:
\begin{equation}
\begin{split}
H=&\sum_{\mathbf{ k},\alpha,\sigma}[\epsilon_{\alpha}(\mathbf{ k})-\mu]c_{\mathbf{ k}\alpha\sigma}^\dag c_{\mathbf{ k}\alpha\sigma}+\sum_{\mathbf{ k},\alpha,\mathbf{ i},\xi,\sigma}(V_{\mathbf{ i}\xi\mathbf{ k}\alpha}d_{\mathbf{ i}\xi\sigma}^\dag c_{\mathbf{ k}\alpha\sigma}\\&+h.c.) +(\varepsilon_d-\mu)\sum_{\mathbf{ i},\xi,\sigma}d_{\mathbf{ i}\xi\sigma}^\dag d_{\mathbf{ i}\xi\sigma}+U\sum_{\mathbf{ i},\xi}n_{\mathbf{ i}\xi\uparrow}n_{\mathbf{ i}\xi\downarrow},
\end{split}
\end{equation}
where $c_{\mathbf{ k}\alpha\sigma}^\dag(c_{\mathbf{ k}\alpha\sigma})$ is the creation (annihilation) operator for a host electron with wave vector $\mathbf{ k}$ and spin $\sigma$ in the valence bands ($\alpha$ = v) or conduction bands ($\alpha$ = c), and $d_{\mathbf{ i}\xi\sigma}^\dag (d_{\mathbf{ i}\xi\sigma})$ is the creation (annihilation) operator for a localized electron at impurity site ${\rm \mathbf{ i}}$ in orbital $\xi$ and spin $\sigma$ with $n_{\mathbf{ i} \xi \sigma}=d_{\mathbf{ i}\xi\sigma}^\dag d_{\mathbf{ i} \xi \sigma}$. Here, $\epsilon_{\alpha}(\mathbf{ k})$ is the band structure of host semiconductor, $\mu$ is the chemical potential, $V_{\mathbf{ i}\xi \mathbf{ k}\alpha}$ denotes the mixing parameter between impurity and host electrons, $\varepsilon_d$ is the impurity 3$d$ orbital energy, and $U$ is the on-site Coulomb repulsion of the impurity. Considering the condition of Hund coupling $J_H\ll U$, $J_H$ is neglected for simplicity in the QMC calculations, and the single-orbital approximation is used to describe the magnetic states of impurities. We calculate magnetic correlations of the impurities using Hirsch-Fye QMC technique with more than 10$^6$ Monte Carlo sweeps and Matsubara time step $\tau$ = 0.25.

\section{Results for (In, Fe)Sb, (In, Mn)Sb and (In, Cr)Sb}

\begin{figure}[!htbp]
  \centering
  \includegraphics[scale=0.45, angle=0]{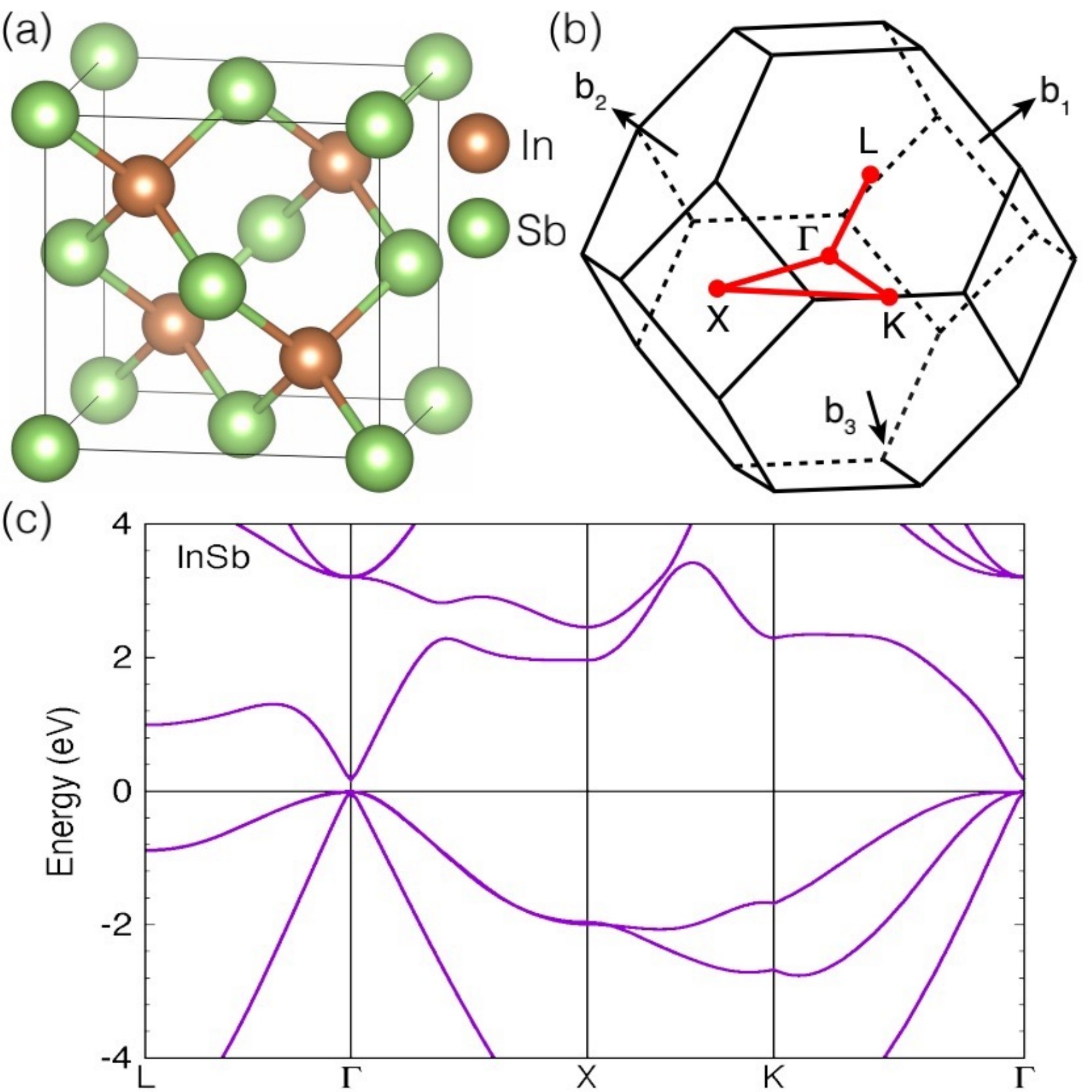}
  \caption{(a) Crystal structure of InSb and its (b) Brillouin zone. (c) Energy bands $\epsilon_\alpha$ of host InSb.}\label{insb_band}
\end{figure}

The parameters $\epsilon_\alpha( \bm{ k})$ and $V_{i \xi \mathbf{k} \alpha}$ are obtained by DFT calculations using the WIEN2k package~\cite{blaha2002wien2k}. To reproduce the experimental narrow band gap of 0.17 eV in InSb, we use the modified Becke-Johnsom (mBJ) exchange potential~\cite{Tran2011}, which has been implemented in the WIEN2k package. The obtained energy band $\epsilon_{\alpha}(\mathbf{k})$ is shown in Fig.~\ref{insb_band}, where InSb has space group F$\overline{4}$3m (No. 216). We obtained a direct gap band $\triangle_g$ = 0.17 eV, which is in good agreement with the experimental value~\cite{Tu2019,Ganesan2008}.

\subsection{Results for (In, Fe)Sb}

The mixing parameter between $\xi$ orbitals of an Fe impurity and InSb host is defined as $V_{i\xi\mathbf{k}\alpha} \equiv \langle\phi_\xi(\mathbf{i})|H|\Psi_\alpha(\mathbf{k})\rangle \equiv \frac{1}{\sqrt{N}}e^{i\mathbf{k}\cdot\mathbf{i}}V_{\xi\alpha}(\mathbf{k})$, which can be expressed as
\begin{equation}
V_{\xi\alpha}(\mathbf{k})=\sum_{o,\mathbf{n}}e^{i\mathbf{k}\cdot(\mathbf{n}-\mathbf{i})}a_{\alpha o}(\mathbf{k}) \langle\phi_\xi(\mathbf{i})|H|\phi_o(\mathbf{n})\rangle,
\end{equation}
where $\phi_\xi(\mathbf{i})$ is the impurity 3$d$ state at site $\mathbf{i}$, and $\Psi_\alpha(\mathbf{k})$ is the host state with wave vector $\mathbf{k}$ and band index $\alpha$, which is expanded by atomic orbitals $\phi_o(\mathbf{n})$ with orbital index $o$ and site index $\mathbf{n}$. Here, $N$ is the total number of host lattice sites and $a_{\alpha o}(\mathbf{k})$ is an expansion coefficient. To obtain the mixing integrals of $\langle\phi_\xi(\mathbf{i})|H|\phi_o(\mathbf{n})\rangle$, we consider a supercell In$_{26}$FeSb$_{27}$, which is comprised of $3\times3\times3$ primitive cells, where each primitive cell consists of an InSb, and an In atom is replaced by an Fe atom. It can be seen that near $\Gamma$ point the orbitals of $d_{xy}$, $d_{xz}$ and $d_{yz}$ contribute more predominantly to the mixing function than the orbitals of $d_{x^2-y^2}$ and $d_{z^2}$, whereas the contributions of $d_{x^2-y^2}$ and $d_{z^2}$ orbitals to the mixing function increase along $\Gamma$ to X or L direction.

For 3$d$ orbitals of transition metal impurities, the reasonable $U$ is estimated as $U$ = 4 eV~\cite{Gu2016,Gu2016a,Gu2008,Ohe2009}. The top of the VB was taken to be zero and the bottom of the CB to be 0.17 eV. $n_\xi$ is defined as $n_\xi=n_{\xi \uparrow}+n_{\xi \downarrow}$. For Fe 3$d$ orbital energy parameter $\varepsilon_d$=-2 eV, a dramatic increase in $n_\xi$ is observed around -0.1 eV for the orbitals $\xi$ = $xy$, $xz$, and $yz$, while for $\varepsilon_d$=-1 eV, sharp increases in $n_\xi$ are observed around -0.6, 0.1 and 0.2 eV for the orbitals $\xi$ = $z^2$ (or $x^2$-$y^2$), $yz$ and $xy$ (or $xz$), respectively. The sharp increase of $n_\xi$ implies the existence of an impurity bound state (IBS) at this energy $\omega_{\rm IBS}$~\cite{Gu2008,Ohe2009,Gu2009,Bulut2007,Tomoda2009}. The operator $M_{\mathbf{i}\xi}^z$ of the $\xi$ orbital at impurity site $\mathbf{i}$ is defined as $M_{\mathbf{i}\xi}^z = n_{\mathbf{i}\xi \uparrow}-n_{\mathbf{i}\xi \downarrow}$. For each $\xi$ orbital, a ferromagnetic (FM) coupling is obtained when the chemical potential $\mu$ is close to the IBS position, and the FM correlations become weaker and eventually disappear when $\mu$ moves away from the IBS. This role of the IBS in determining the strength of FM correlations between impurities is consistent with the Hartree-Fock and QMC results of various diluted ferromagnetic semiconductors~\cite{Gu2008,Ohe2009,Gu2009,Bulut2007,Tomoda2009}.

Recent experiments showed high Curie temperature Tc in n-type semiconductor (In,Fe)Sb~\cite{Kudrin2017,Tu2018,Tu2019}. In order to reproduce the FM coupling between Fe impurities in semiconductor InSb with n-type carriers, the reasonable 3$d$ orbital energy parameter is $\varepsilon_d$ = -1 eV. For n-type carriers, the chemical potential is about $\mu$ = 0.17 eV, i.e. near the bottom of conduction band. The FM coupling between Fe impurities is obtained for $\varepsilon_d$ = -1 eV with $\mu$ = 0.17 eV, while no stable magnetic coupling between Fe impurities is obtained for $\varepsilon_d$ = -2 eV with $\mu$ = 0.17 eV. Thus, as a reasonable estimation, we take the parameter $\varepsilon_d$ = -1 eV for 3$d$ orbitals in our following QMC calculations.

Long-range FM coupling up to approximately 8 \AA \ (1.4$a_0$) is obtained for the orbitals $\xi$ = $yz$ for the p-type case with $\mu$ = 0 eV. Long-range FM coupling up to approximately 13 \AA \ (2$a_0$) is obtained for the orbitals $\xi$ = $xz$ as well. A comparison shows that the magnitude of FM coupling $\langle M_{1\xi}^zM_{2\xi}^z \rangle$ in the $n$-type case is slightly larger than that in the $p$-type case, which is consistent with the high Tc observed in the n-type semiconductor (In,Fe)Sb in experiments~\cite{Kudrin2017,Tu2018,Tu2019}. In addition, our calculation predicts that it is possible to obtain high Tc in semiconductor (In,Fe)Sb with p-type carriers.

\subsection{Results for (In, Mn)Sb}

We made similar calculations for Mn-doped InSb. It is noted that the mixing parameters are similar to those of Fe-doped InSb. Figure shows the occupation number $\langle n_\xi \rangle$ of $\xi$ orbital of Mn impurity in InSb versus chemical potential $\mu$ at temperature 360 K. Sharp increases in $n_\xi$, which imply the position of IBS $\omega_{\rm IBS}$, were observed around -0.5 eV for $x^2-y^2$ and $z^2$ orbitals,  and around 0 eV for $xy$, $xz$ and $yz$ orbitals, respectively. Figure shows the magnetic correlation $\langle M_{1\xi}^zM_{2\xi}^z \rangle$ between $\xi$ orbitals of two Mn impurities as a function of $\mu$, with fixed distance $R_{12}$ as the nearest neighbor. The role of the IBS in determining the strength of FM correlations between impurities is the same as that discussed for Fe-doped InSb. Figure shows the distance $R_{12}$ dependent magnetic correlation $\langle M_{1\xi}^zM_{2\xi}^z \rangle$ between $\xi$ orbitals of two Mn impurities for $p$- and $n$-type cases. For p-type carriers with $\mu$ = 0 eV, the same value used for p-type (In,Fe)Sb, a long-range FM coupling up to approximately 10 \AA \ was obtained for $\xi$ = $yz$ orbital, which is longer than 8 \AA \ obtained for Fe-doped InSb with $p$-type carriers, while a relatively short-range FM coupling was obtained for the $\xi$ = $xy$ and $xz$ orbitals for $p$-type case. However, for $n$-type carriers with $\mu$ = 0.15 eV, a weak FM coupling was obtained. Comparing with Figs., we find that the FM couplings in (In, Fe)Sb and (In, Mn)Sb are similar. In experiments, high Tc was obtained in n-type (In,Fe)Sb~\cite{Kudrin2017,Tu2018,Tu2019}, but low Tc was obtained in (In,Mn)Sb~\cite{Ganesan2008}. In Sec. VI, we will discuss our understandings in a general picture. In addition, our results predict that a weak FM can exist in semiconductor (In,Mn)Sb with n-type carriers.

\subsection{Result for (In, Cr)Sb}

We made similar calculations for Cr-doped InSb that has not yet been attempted in experiments. It is clear that the mixing parameters are very similar for Fe, Mn, Cr-doped InSb. The chemical potential $\mu$ dependent occupation number $\langle n_\xi \rangle$ of $\xi$ of a Cr impurity and the magnetic correlation $\langle M_{1\xi}^zM_{2\xi}^z \rangle$ between $\xi$ orbitals of two Cr impurities of the nearest neighbors are shown in Fig.. Comparing with Figs., the similar behaviors of IBS and magnetic correlation are obtained for Fe, Mn, Cr-doped InSb. The distance $R_{12}$ dependent magnetic correlation $\langle M_{1\xi}^zM_{2\xi}^z \rangle$ between $\xi$ orbitals of two Cr impurities with p- and n-type cases is shown in Fig..
Comparing with Figs., the similar magnetic correlations are obtained for Fe, Mn, Cr-doped InSb. Thus, our calculations predict the p- and n-type ferromagnetic semiconductors Cr-doped InSb that possibly have high Tc.

\section{Results for Fe, Mn and Cr-doped InAs}

\begin{figure}[!htbp]
  \centering
  \includegraphics[scale=0.75, angle=0]{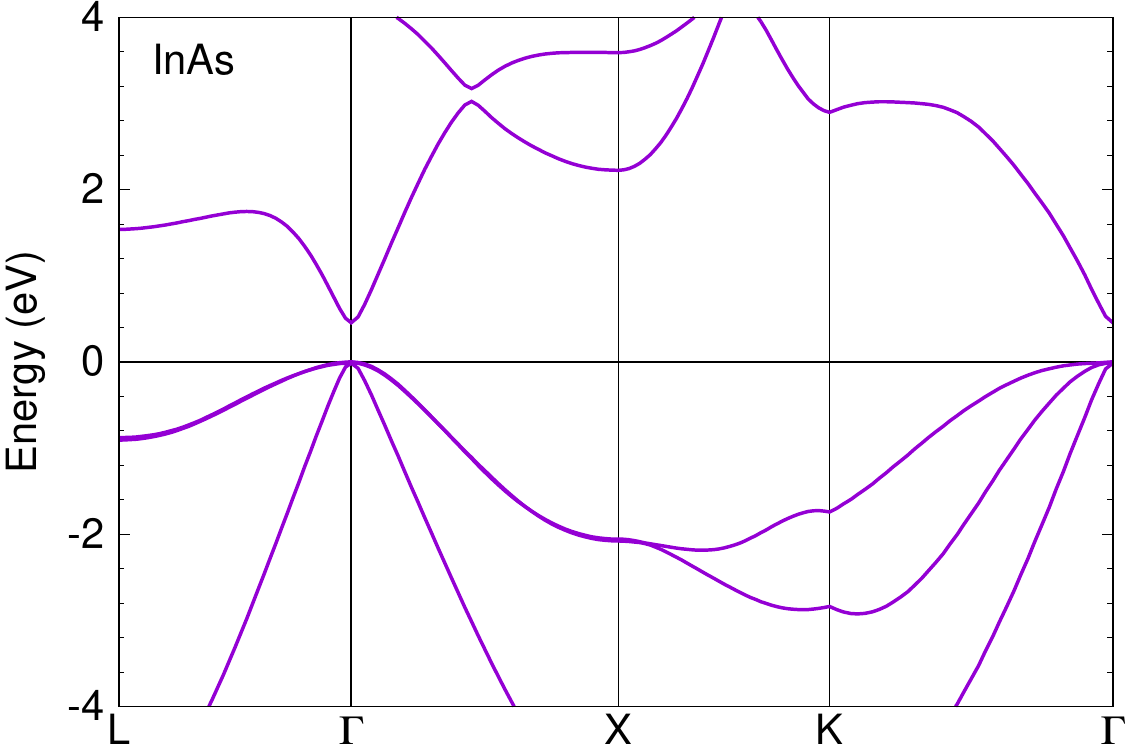}
  \caption{Energy bands $\epsilon_\alpha$ of host InAs, which has the same space group as InSb. A direct band gap of 0.42 eV was obtained by DFT calculations, which agrees well with the experimental value.}\label{inas_band}
\end{figure}

InAs has the same space group of F$\overline{4}$3m (No. 216) as InSb, but has a smaller lattice constant of 6.05 \AA. By DFT method with the modified Becke-Johnsom (mBJ) exchange potential, we calculate the band structure of InAs as shown in Fig.~\ref{inas_band}. The obtained band gap of 0.42 eV is in a good agreement with the experimental value~\cite{Hai2012,Schallenberg2006}. 

\subsection{Results for (In, Fe)As}

For Fe-doped InAs, chemical potential $\mu$ dependent occupation number $\langle n_\xi \rangle$ of $\xi$ orbitals of an Fe impurity and magnetic correlation $\langle M_{1\xi}^zM_{2\xi}^z \rangle$ between $\xi$ orbitals of two Fe impurities of the first-nearest neighbor is shown in Fig.. It is shown that the IBS appears at about 0.1 eV for orbitals $xy$, $xz$ and $yz$. The FM correlations between orbitals $xy$, $xz$, and $yz$ were obtained when chemical potential $\mu$ is close to the position of IBS. The distance $R_{12}$ dependent magnetic correlation $\langle M_{1\xi}^zM_{2\xi}^z \rangle$ between $\xi$ orbitals of two Fe impurities for the p-type case with chemical potential $\mu$ = 0.1 eV is shown in Fig..
The long range FM coupling between $yz$ orbitals of Fe impurities is obtained. Comparing with Figs., it is shown that the FM couplings in (In, Fe)Sb and (In, Fe)As are similar. In experiments, high Tc was obtained in n-type (In,Fe)Sb~\cite{Kudrin2017,Tu2018,Tu2019}, but low Tc was obtained in p-type (In,Fe)As~\cite{Hai2012,Hai2012a}. In Sec. VI, we give a general picture for this observation.

\subsection{Results for (In, Mn)As}

For Mn-doped InAs, chemical potential $\mu$ dependent occupation number $\langle n_\xi \rangle$of $\xi$ orbitals of a Mn impurity and magnetic correlation $\langle M_{1\xi}^zM_{2\xi}^z \rangle$ between $\xi$ orbitals of two Mn impurities of the first-nearest neighbor is shown in Fig.. It can be observed that the IBS appears at about 0.1$\sim$0.2 eV for orbitals $xy$, $xz$ and $yz$. The FM correlations between orbitals $xy$, $xz$, and $yz$ were obtained when chemical potential $\mu$ is close to the position of IBS. The distance R12 dependent magnetic correlation $\langle M_{1\xi}^zM_{2\xi}^z \rangle$ between $\xi$ orbitals of two Mn impurities for the p-type case with chemical potential $\mu$ = 0.1 eV is shown in Fig.. The long range FM coupling between $yz$ orbitals of Mn impurities is obtained. Comparing with Figs., we uncover that the FM couplings in (In,Fe)As and (In,Mn)As are similar. In experiments, low and similar Tc were obtained in n-type (In,Fe)As~\cite{Hai2012,Hai2012a} and p-type (In,Mn)As~\cite{Schallenberg2006}. The detail discussions will be presented in Sec. VI.

\subsection{Results for (In, Cr)As}

For Cr-doped InAs, the chemical potential $\mu$ dependent occupation number $\langle n_\xi \rangle$ of a Cr impurity and the magnetic correlation $\langle M_{1\xi}^zM_{2\xi}^z \rangle$ between $\xi$ orbitals of
two Cr impurities of the first nearest neighbor are shown in Fig.. Comparing with Figs., similar behaviors of IBS and magnetic correlation are obtained for Fe, Mn, Cr-doped InAs. The distance $R_{12}$ dependent magnetic correlation $\langle M_{1\xi}^zM_{2\xi}^z \rangle$ between $\xi$ orbitals of two Cr impurities with p-type carriers is shown in Fig.. Comparing with Figs., similar magnetic correlations are also obtained for Fe, Mn, Cr-doped InAs. Thus, our calculations predict a ferromagnetic semiconductor Cr-doped InAs with p-type carriers.

\section{Results for Fe, Mn and Cr doped GaSb}

\begin{figure}[!htbp]
  \centering
  \includegraphics[scale=0.75, angle=0]{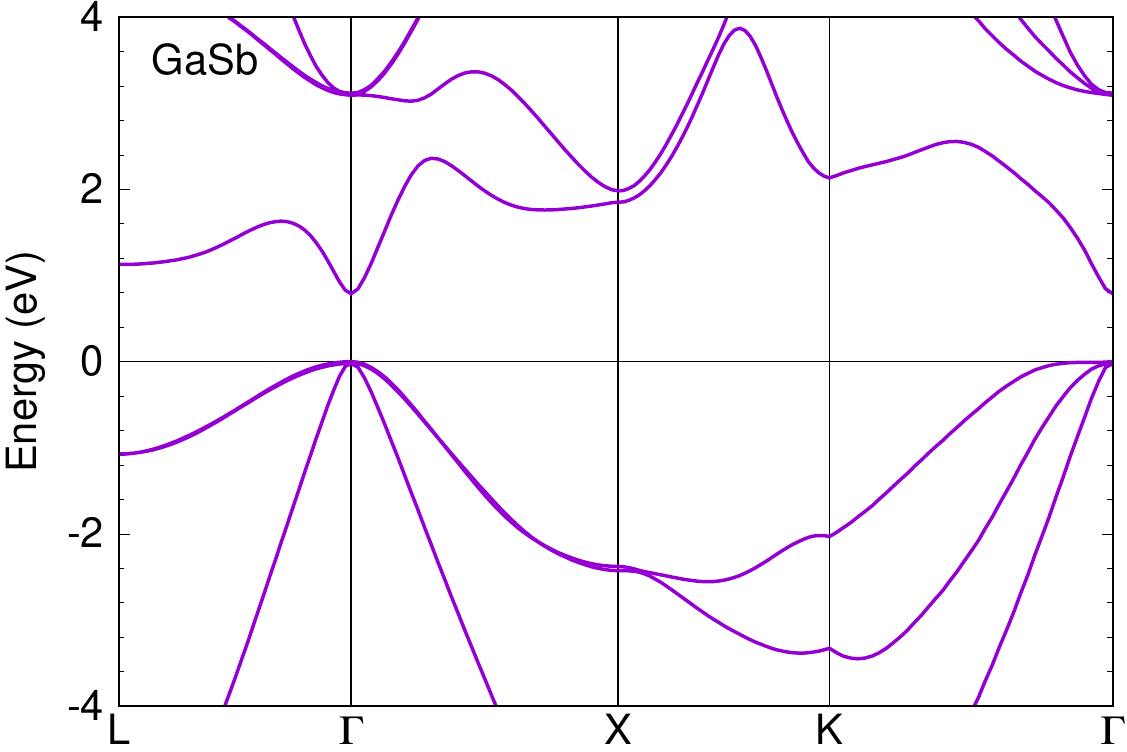}
  \caption{Energy bands $\epsilon_\alpha$ of host GaSb, which has the same space group as InSb. A direct band gap of 0.75 eV was obtained by DFT calculations, which agrees well with the experimental value.}\label{gasb_band}
\end{figure}

GaSb has the same space group of F$\overline{4}$3m (No. 216) as InSb, but has a smaller lattice constant of 6.09 \AA. By DFT method with the modified Becke-Johnsom (mBJ) exchange potential, we calculate the band structure of GaSb as shown in Fig.~\ref{gasb_band}. The obtained band gap of 0.75 eV is in a good agreement with the experimental value~\cite{Tu2016,Abe2000}.

\subsection{Results for (Ga, Fe)Sb}

For Fe-doped GaSb, chemical potential $\mu$ dependent occupation number $\langle n_\xi \rangle$ of $\xi$ orbitals of an Fe impurity and magnetic correlation $\langle M_{1\xi}^zM_{2\xi}^z \rangle$ between $\xi$ orbitals of two Fe impurities of the first-nearest neighbor is shown in Fig.. It is shown that the IBS appears at about 0.1 eV for orbital $yz$, and about 0.25 eV for $xz$ and $xy$ orbitals. The FM correlations between orbitals $xy$, $xz$, and $yz$ were obtained when chemical potential $\mu$ is close to the position of IBS. The distance $R_{12}$ dependent magnetic correlation $\langle M_{1\xi}^zM_{2\xi}^z \rangle$ between $\xi$ orbitals of two Fe impurities for the p-type case with chemical potential $\mu$ = 0 eV is shown in Fig.. The long range FM coupling between $yz$ orbitals of Fe impurities is obtained,
which is consistent with the high Tc observed in the p-type semiconductor (Ga,Fe)Sb in experiments~\cite{Tu2016,Goel2019,Goel2019a,Takiguchi2019}.

\subsection{Results for (Ga, Mn)Sb}

For Mn-doped GaSb, chemical potential $\mu$ dependent occupation number $\langle n_\xi \rangle$ of $\xi$ orbitals of a Mn impurity and magnetic correlation $\langle M_{1\xi}^zM_{2\xi}^z \rangle$ between $\xi$ orbitals of two Mn impurities of the first-nearest neighboris shown in Fig.. It is shown that the IBS appears at a bout 0.1 eV for orbitals $xy$, $xz$ and $yz$. The FM correlations between orbitals $xy$, $xz$, and $yz$ were obtained when chemical potential $\mu$ is close to the position of IBS. The distance $R_{12}$ dependent magnetic correlation $\langle M_{1\xi}^zM_{2\xi}^z \rangle$ between $\xi$ orbitals of two Mn impurities for the p-type case with chemical potential $\mu$ = 0.1 eV is shown in Fig.. The long range FM coupling between $yz$ orbitals of Mn impurities is obtained. Comparing with Figs., one may find that the FM couplings in (Ga, Fe)Sb and (Ga,Mn)Sb are similar. In experiments, high Tc was obtained in p-type (Ga,Fe)Sb\cite{Tu2016,Goel2019,Goel2019a,Takiguchi2019}, but low Tc was obtained in p-type (Ga,Mn)Sb~\cite{Abe2000}. To resolve this puzzling situation, in Sec. VI we will discuss this issue in a general picture.

\subsection{Results for (Ga, Cr)Sb}

For Cr-doped GaSb, the chemical potential $\mu$ dependent occupation number $\langle n_\xi \rangle$ of a Cr impurity and the magnetic correlation $\langle M_{1\xi}^zM_{2\xi}^z \rangle$ between $\xi$ orbitals of
two Cr impurities of the first nearest neighbor are shown in Fig.. Comparing with Figs., similar behaviors of IBS and magnetic correlation are obtained for Fe, Mn, Cr-doped GaSb. The distance $R_{12}$ dependent magnetic correlation $\langle M_{1\xi}^zM_{2\xi}^z \rangle$ between $\xi$ orbitals of two Cr impurities with p-type carriers is shown in Fig.. Comparing with Figs., similar magnetic correlations are obtained for Fe, Mn, Cr-doped GaSb.
Thus, our calculations predict the p-type ferromagnetic semiconductor Cr-doped GaSb could have high Tc. 

\section{Discussion}

\begin{table}[!!htbp]
  \caption{The DFT+QMC calculation results of the maximum value of $\langle M_{1}M_{2} \rangle$ between two impurities with the first nearest neighbor ($n.n.$) for Fe, Mn, Cr-doped InSb, InAs, and GaSb.}
  \label{tab:1}
  \begin{tabular}{c|c|c|c}
		\hline
		\multirow{3}{2.1cm}{The maximum $\langle M_{1}M_{2} \rangle$ at the first $n.n.$}   &
		\multicolumn{3}{c}{Host semiconductors} \\
		 \cline{2-4} 
		 &InSb &InAs   &GaSb  \\
		 &(gap=0.17eV)  &(gap=0.42eV) &(gap=0.75eV) \\
		\hline 
		\multirow{2}*{Fe}    &0.07 (p-type)  &0.14 (p-type)   &0.1 (p-type) \\
		&0.09 (n-type) & &\\
		\hline
               \multirow{2}*{Mn}    &0.14 (p-type)  &0.13 (p-type)   &0.13 (p-type) \\
		&0.07 (n-type) & &\\
		\hline
               \multirow{2}*{Cr}    &0.13 (p-type)  &0.12 (p-type)   &0.12 (p-type) \\
		&0.08 (n-type) & &\\
		\hline
	\end{tabular}
\end{table}

To understand the mechanism of ferromagnetism in these ferromagnetic semiconductors, we list the results of
the maximum $\langle M_{1\xi}^zM_{2\xi}^z \rangle$ between two impurities with the first nearest neighbor (n.n.) in Table~\ref{tab:1}. The value of the maximum $\langle M_{1\xi}^zM_{2\xi}^z \rangle$ for impurities Fe, Mn and Cr is respectively about 0.1, 0.13 and 0.12 for p-type semiconductor GaSb, about 0.14, 0.13 and 0.12 for p-type semiconductor InAs, and about 0.09, 0.07 and 0.08 for n-type semiconductor InSb. It is noted that different values of the maximum $\langle M_{1\xi}^zM_{2\xi}^z \rangle$ are obtained for different impurities in p-type InSb. The maximum $\langle M_{1\xi}^zM_{2\xi}^z \rangle$ for impurities Fe, Mn and Cr is respectively about 0.07, 0.14 and 0.13 for p-type semiconductor InSb. In general, the results suggest that for the same host semiconductors, impurities Fe, Mn, and Cr have quite similar magnetic correlation $\langle M_{1\xi}^zM_{2\xi}^z \rangle$. The
results in Table~\ref{tab:1} are calculated based on the Anderson impurity model in the diluted impurity limit. Based on these results, we argue that high (low) Tc in these ferromagnetic semiconductors in experiments may come from high (low) impurity concentrations, rather than from the impurity-impurity magnetic correlations themselves.

\begin{table*}[!!htbp]
  \caption{The experimental results of Curie temperature, impurity concentration, carrier type, and band gap for some ferromagnetic semiconductors.}
  \label{tab:2}
  \begin{tabular}{c|c|c|c|c|c}
		\hline
		Ferromagnetic   &  Curie temperature, & Impurity & Carrier types & Band gap & Reference of \\ 
		semiconductors & Tc (K)                    & concentration, N$_m$  &  & (eV)        & experiments (years)\\
		 \hline 
		 (Ga,Fe)Sb        &340         &25$\%$       &p-type      &0.75    &Ref.~\mycite{Tu2016} (2016)\\
		 \hline
		 (Ga,Mn)Sb        &25          &2.3$\%$       &p-type      &0.75         &Ref.~\mycite{Abe2000} (2000)\\
		 \hline
		 (In,Fe)Sb        &385          &35$\%$       &n-type      &0.17         &Ref.~\mycite{Tu2019} (2019)\\
		 \hline
		 (In,Mn)Sb        &10          &4$\%$       &p-type      &0.17         &Ref.~\mycite{Ganesan2008} (2008)\\
		 \hline
		 (In,Fe)As        &70        &8$\%$       &n-type      &0.42         &Ref.~\mycite{Hai2012} (2012)\\
		 \hline
		 (In,Mn)As        &90      &10$\%$       &p-type      &0.42       &Ref.~\mycite{Schallenberg2006} (2006)\\
		 \hline
		(Ba,K)(Zn,Mn)$_2$As$_2$   &230    &15$\%$       &p-type      &0.2     &Ref.~\mycite{Zhao2014} (2014)\\
		 \hline
		Ba(Zn,Co)$_2$As$_2$        &45           &4$\%$       &n-type      &0.2         &Ref.~\mycite{Guo2019} (2019)\\
		 \hline
		 Li(Zn,Mn)As        &50       &10$\%$       &p-type      &1.6         &Ref.~\mycite{Deng2011} (2011)\\
		 \hline
		 Li(Zn,Mn)P        &34         &10$\%$       &p-type      &2.0         &Ref.~\mycite{Deng2013} (2013)\\
		 \hline
		 Li(Cd,Mn)P        &45         &10$\%$       &p-type      &1.3         &Ref.~\mycite{Han2019} (2019)\\
		 \hline
		 (Ga,Mn)As        &200        &16$\%$       &p-type      &1.4        &Ref.~\mycite{Chen2011} (2011)\\
		 \hline
	\end{tabular}
\end{table*}

To further check our argument, in Table~\ref{tab:2} we list the experimental results of Curie temperature Tc and the impurity concentrations N$_m$ in some ferromagnetic semiconductors. From Table~\ref{tab:2}, it is observed that for the same host semiconductors with different impurities, Tc is naively proportional to the impurity concentration N$_m$. For example, Tc = 340K, N$_m$=25$\%$ for Fe-doped GaSb, and Tc = 25K, N$_m$ = 2.3$\%$ for Mn-doped GaSb. More examples are available in Table~\ref{tab:2}, such as Tc = 70 K, N$_m$ = 8$\%$ for Fe-doped InAs, and Tc = 90 K, N$_m$ = 10$\%$ for Mn-doped InAs; Tc = 230K, N$_m$ = 15$\%$ for Mn-doped BaZn$_2$As$_2$, and Tc = 45K, N$_m$ = 4$\%$ for Co-doped BaZn$_2$As$_2$. Certainly, this relation is not numerically exact, and the deviation from this relation exists. For example, Tc = 385K, N$_m$ = 35$\%$ for Fe-doped InSb, and Tc = 10K, N$_m$ = 4$\%$ for Mn-doped InSb. In order to increase Tc in diluted ferromagnetic semiconductors, our results suggest that a primary strategy is to increase the impurity concentrations. For this purpose, it is crucial to choose proper impurities and semiconductor hosts to avoid valence state mismatch during the magnetic doping. 

Our numerical results show that the reason that (Ga,Fe)Sb and (In,Fe)Sb have high Tc in the experiments~\cite{Tu2016, Tu2019} is from high impurity concentrations of Fe impurities. The Fe impurities in GaSb and InSb do not introduce additional carriers~\cite{Tu2016, Tu2019}, which means that there is no valence mismatch between Fe$^{3+}$ and Ga$^{3+}$ and In$^{3+}$, and Fe impurities can be doped to vary high concentrations (25$\%$ to 35$\%$ in Table~\ref{tab:2}). In contrast, Mn impurities in GaSb and InSb is Mn$^{2+}$, and cannot be doped in a high concentration due to valence mismatch. 

Therefore, the origin of high Tc in (Ga,Fe)Sb and (In,Fe)Sb is not due to the carrier induced mechanism because Fe$^{3+}$ does not introduce carriers. 

The calculations show that ferromagnetism appears in both p- and n-type semiconductors with narrow band gaps, and in only p-type semiconductors with wide band gaps, which is consistent with our previous theory~\cite{Gu2016,Gu2016a,Gu2019} and experimental observations in most of ferromagnetic semiconductors. 

\section{Curie temperature by mean field theory}

We estimate the Curie temperature Tc from a simple Weiss mean-field formula:
\begin{equation}\label{Curie}
{\rm Tc} = \frac{2}{3k_B}S(S+1)\sum_{i}z_iJ_i,
\end{equation}
where $z_i$ is the coordination number and $J_i$ are the exchange couplings of the $i$-th nearest neighbors.

To obtain the exchange couplings, as an approximation, we map the ferromagnetic correlation between impurities onto the isotropic Heisenberg model for two particles with five $d$ orbitals
\begin{equation}
H=-J\sum_{\zeta,\eta=d}\mathbf{S}_{1\zeta}\cdot \mathbf{S}_{2\eta},
\end{equation}
where $\mathbf{S}_{1\zeta}$ represents the spin-1/2 operator of $\zeta$ (= $d_{xy}$, $d_{xz}$, $d_{yz}$, $d_{x^2-y^2}$ or $d_{z^2}$) orbital. At finite temperature, by defining $\beta$ = 1/$k_B$T with $k_B$ the Boltzmann constant, we write down the orbital-orbital correlation as
\begin{equation}
\langle S_{1m}^zS_{2n}^z \rangle = {\rm Tr}(S_{1m}^zS_{2n}^ze^{-\beta H})/{\rm Tr}(e^{-\beta H}),
\end{equation}
where $S_i^z$ represents the z component of spin $\mathbf{S}_i$.
Considering 
\begin{equation}
\begin{split}
\mathbf{S}_{1\zeta}\cdot \mathbf{S}_{2\eta} &= \frac{1}{2}[(\mathbf{S}_{1\zeta}+\mathbf{S}_{2\eta})^2-\mathbf{S}_{1\zeta}^2 - \mathbf{S}_{2\eta}^2] \\ 
&=\frac{1}{2}[S_{\zeta\eta}(S_{\zeta\eta}+1)-2s(s+1)],
\end{split}
\end{equation}
where $s$ = 1/2, $S_{\zeta\eta}$ = 0, 1, and the trace could be taken as summation:
\begin{equation}
\begin{split}
{\rm Tr}(\cdots) &=\prod_{\zeta, \eta}[\sum_{S_{\zeta\eta}=0}^1\sum_{S_{\zeta\eta}^z=-S_{\zeta\eta}}^{S_{\zeta\eta}}](\cdots) \\
& = \prod_{\zeta, \eta}[\sum_{S_{1\zeta}^z=-1/2}^{1/2}\sum_{S_{2\eta}^z=-1/2}^{1/2}](\cdots),
\end{split}
\end{equation}
\begin{equation}
\begin{split}
{\rm Tr}(e^{-\beta H}) &=[\sum_{S_{mn}=0}^1\sum_{S_{mn}^z=-S_{mn}}^{S_{mn}}e^{\beta J S_{1m}S_{2n}}]\times \\
&\prod_{\zeta, \eta \neq m, n}[\sum_{S_{\zeta\eta}=0}^{1}\sum_{S_{\zeta\eta}^z=-S_{\zeta\eta}}^{S_{\zeta\eta}}e^{\beta J S_{1\zeta}S_{2\eta}}],
\end{split}
\end{equation}
we have 
\begin{equation}
\langle S_{1m}^zS_{2n}^z \rangle= \frac{1}{4}\cdot \frac{1-e^{-\beta J}}{3+e^{-\beta J}}.
\end{equation}
We define the average of impurity-impurity correlation $\langle M_1^zM_2^z \rangle = \sum_{\zeta, \eta=d}\langle M_{1\zeta}^zM_{2\eta}^z \rangle/(2S)^2$, where spin $S = \langle n \rangle/2 = \sum_\xi \langle n_\xi \rangle/2$, and the results of $\langle M_1^zM_2^z \rangle$ extracted from our QMC calculation results are listed in Table ~\cite{}. The unit of $\langle M_1^zM_2^z \rangle$ is 1$^2$ = (2$s$)$^2$ = 4($\mu_B$)$^2$ and $\mu_B$ is Bohr magneton. To be consistent with our QMC calculation results, whose unit of $\langle M_1^zM_2^z \rangle$ is $\mu_B^2$, the average of impurity-impurity correlation can be obtained as $\langle M_1^zM_2^z \rangle = 4\langle S_{1m}^zS_{2n}^z \rangle$. Thus, for a given temperature if we have the average of impurity-impurity correlation
$\langle M_1^zM_2^z \rangle$, we can deduce an effective exchange $J$ between two impurities.

\begin{table}[!!htbp]
  \caption{The $\langle M_1^zM_2^z \rangle$ and the effective exchange $J_i$ extracted from DFT+QMC calculations between two Fe impurities with different nearest neighbor ($n.n.$) for n-type Fe-doped InSb with spin $S$ = 5/2. We have only plotted $\langle M_{1\zeta}^zM_{2\eta}^z \rangle$ with $\zeta = \eta = d_{xz}, d_{yz}$, because the orbital-orbital correlation $\langle M_{1\zeta}^zM_{2\eta}^z \rangle$ for other $\zeta$ and $\eta$ orbitals are calculated to be very small and negligible.}
  \label{tab:3}
  \begin{tabular}{c|c|c|c|c}
		\hline
		 the $i$-th $n.n.$    &first    &second     &third     &fourth \\
		\hline 
		$\langle M_1^zM_2^z \rangle$ (10$^{-2}$ $\mu_B^2$)     &0.467 &0.231 &0.155 &0.124\\
		\hline
		$J_i/k_B$ (K)  &6.757 &3.334 &2.235 &1.788\\
		\hline
	\end{tabular}
\end{table}

\begin{figure*}[!htbp]
  \centering
  \includegraphics[scale=0.6,angle=0]{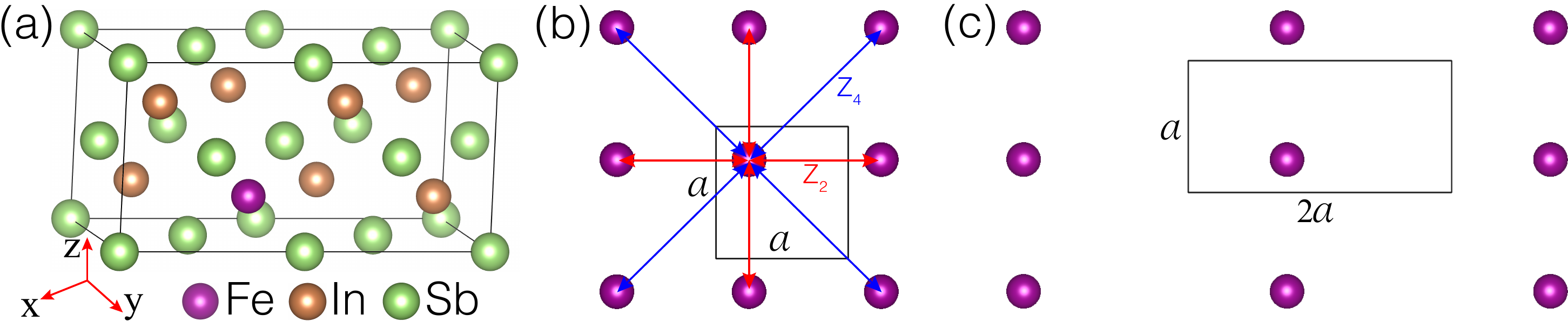}
  \caption{(a) 12.5$\%$ doping concentration for Fe-doped InSb in a supercell of InSb containing eight In atoms, and the views from (b) [100] and (c) [001] directions, where we have only plotted the magnetic atoms (Fe atoms)  and the supercell is indicated.}\label{doping_0.125}
\end{figure*}

In the following, we shall take the n-type (In,Fe)Sb as an example to investigate the impurity concentration dependent Curie temperature. The total spin for n-type (In,Fe)Sb is $S=\langle n \rangle/2 = 5/2$. The impurity-impurity correlation $\langle M_1^zM_2^z \rangle$ between two Fe impurities as a function of the distance between two Fe impurities and the effective exchange $J_i$ of the $i$-th $n.n.$ are listed in Table~\ref{tab:3}, where we only consider four kinds of $n.n.$ hoppings with the $n.n.$ $R_1$ = (0.5, 0, 0.5)a, the next $n.n.$ $R_2$ = (1, 0, 0)a, the third $n.n.$ $R_3$ = (1, 0.5, 0.5)a and the fourth $n.n.$ $R_4$ = (1, 1, 0)a.

\begin{table}[!!htbp]
  \caption{The Curie temperatures corresponding to three different doping concentration of 12.5$\%$, 25$\%$ and 37.5$\%$ for n-type Fe-doped InSb.}
  \label{tab:4}
  \begin{tabular}{c|c|c|c}
		\hline
		 doping concentration    &12.5$\%$    &25$\%$    &37.5$\%$ \\
		\hline 
		Tc (K)  &120 &245 &379\\
		\hline
	\end{tabular}
\end{table}

To simulate different doping concentrations, for simplicity, we consider three different doping concentrations (Nm) of 12.5$\%$, 25$\%$ and 37.5$\%$ in a supercell of InSb containing eight In atoms (two unitcells). In this case the three different doping concentrations correspond to one, two and three Fe atom impurities, respectively. If we consider larger supercells, these three concentrations will hold more doping configurations, which are too complicated situations and we shall not study here. Let us first investigate the 12.5$\%$ doping concentration, i.e. one In atom was replaced by a Fe impurity in the InSb supercell as shown in Fig.~\ref{doping_0.125}(a). In this case, from Figs.~\ref{doping_0.125}(b) and (c), one may observed that the Fe impurities form a two-dimensional square lattice with the lattice constant a ($R_2$), so the coordination numbers $z_i$ = $\{$0, 4, 0, 4$\}$ with $i$ = $\{$1, 2, 3, 4$\}$. Thus, combining Eq. (\ref{Curie}) and Table~\ref{tab:3} we can estimate the Curie temperature to be 120 K for 12.5$\%$ n-type Fe-doped InSb. For 25$\%$ doping concentration, i.e. two Fe impurities in the InSb supercell containing eight In atoms, there are 28 doping configurations, among which only three are inequivalent with the probability of 5/7, 1/7 and 1/7, respectively, giving rise to a Curie temperature by Eq. (\ref{Curie}) of 250.5, 241.8 and 223.8 K, respectively. Thus, the average Curie temperature can be estimated to be 245 K for 25$\%$ n-type Fe-doped InSb. For 37.5$\%$ doping concentration, i.e. three In atoms were replaced by three Fe impurities in the InSb supercell containing eight In atoms. There are 56 doping configurations, among which only four are inequivalent. The probability of the four different doping methods is 2/7, 2/7, 1/7 and 2/7, respectively. They give rise to a Curie temperature by Eq. (\ref{Curie}) of 399.3, 375.7, 375.7 and 363.7 K, respectively. Thus, the average Curie temperature can be estimated to be 379 K for 37.5$\%$ n-type Fe-doped InSb. These results are summarized in Table~\ref{tab:4}. Our estimated Curie temperature of 379 K for 37.5$\%$ n-type Fe-doped InSb is comparable with the experimental Curie temperature of 385 K for 35$\%$ n-type Fe-doped InSb in Table~\ref{tab:2}.

\begin{table}[!!htbp]
  \caption{The Curie temperatures corresponding to three different doping concentration of 12.5$\%$, 25$\%$ and 37.5$\%$ for p-type Cr-doped InSb, InAs and GaSb.}
  \label{tab:5}
  \begin{tabular}{c|c|c|c|c}
		\hline
		 &doping concentration    &12.5$\%$    &25$\%$    &37.5$\%$ \\
		\hline 
		\multirow{3}*{Tc (K)} &(In,Cr)Sb  &62 &214 &406 \\
		\cline{2-5}
		                               &(In,Cr)As  &17 &129 &264 \\
		\cline{2-5}
		                               &(Ga,Cr)Sb  &242 &534 &851 \\
		\hline
	\end{tabular}
\end{table}

By using the above method and combining the QMC results, we estimate the Curie temperatures corresponding to three different doping concentration of 12.5$\%$, 25$\%$ and 37.5$\%$ for p-type Cr-doped InSb, InAs and GaSb as listed in Table~\ref{tab:5}. From Table~\ref{tab:5}, we find that p-type Cr-doped InSb, InAs and GaSb ferromagnetic semiconductors may have high Tc with a high doping concentration, especially for p-type Cr-doped GaSb.

\section{Summary}
By the combined method of density functional theory and quantum Monte Carlo, we have systematically studied the ferromagnetism of Fe, Mn, Cr-doped GaSb, InSb, and InAs magnetic semiconductors. In the diluted impurity limit, our calculations show that the impurities Fe, Mn, and Cr have similar magnetic correlations in the same host semiconductors. We predict that ferromagnetic semiconductors of Cr-doped InSb,
InAs, and GaSb may have possibly high Tc. More importantly, our results imply that high (low) Tc obtained
in these experiments mainly come from high (low) impurity concentrations rather than the magnetic correlations
between impurities. In addition, our results show that the origin of high Tc in (Ga,Fe)Sb and (In,Fe)Sb is not due to the carrier induced mechanism because Fe$^{3+}$ does not introduce carriers.  In order to increase the Curie temperature in diluted ferromagnetic semiconductors, our results suggest that an alternative tactics is to increase the impurity concentrations, such as by choosing proper impurities and host semiconductors to avoid valence state mismatch during the magnetic doping. 

\section{Acknowledgement}
The authors thank valuable discussions with Prof. M. Tanaka on related experiments. This work is supported in part by the National Key R\&D Program of China (Grant No. 2018YFA0305800), the Strategic Priority Research Program of the Chinese Academy of Sciences (Grant No. XDB28000000), the National Natural Science Foundation of China (Grant No.11834014), and Beijing Municipal Science and Technology Commission (Grant No. Z191100007219013). B.G. is also supported by the National Natural Science Foundation of China (Grant No. Y81Z01A1A9), the Chinese Academy of Sciences (Grant No. Y929013EA2), the University of Chinese Academy of Sciences (Grant No. 110200M208), the Strategic Priority Research Program of Chinese Academy of Sciences (Grant No. XDB33000000), and the Beijing Natural Science Foundation (Grant No. Z190011).

\bibliographystyle{apsrev4-1}
\bibliography{ref}

\end{document}